# Extreme Polarization of the Optical Gap and High-Energy Exciton Landscape in CrSBr

Sayantan Patra[1], Sourabh Jain[1], Bhumika Chauhan[1], Marie-Christin Heißenbüttel[3], Abhisek Saidarsan[1], Ranjuna M. K.[1], Kseniia Mosina[2], Zdeněk Sofer[2], Michael Rohlfing[3], Thorsten Deilmann[3], Ashish Arora[1,*]

[1]*Department of Physics, Indian Institute of Science Education and Research, Dr. Homi Bhabha Road, 411008 Pune, India*
[2]*Department of Inorganic Chemistry, University of Chemistry and Technology, Prague 6, Technicka 5, 16628, Prague, Czech Republic*
[3] *Institute of Solid-State Theory, University of Münster, D-48149 Münster, Germany*

Email: ashish.arora@iiserpune.ac.in

We reveal a strongly anisotropic excitonic landscape in monolayer and bulk-like CrSBr using optical absorption spectroscopy and $GW$-Bethe-Salpeter equation *ab initio* calculations. The direct absorptive determination of the lowest bright optical onsets i.e. $X_0^a$ and $X_0^b$ excitons for the two in-plane polarization eigenaxes yield an in-plane optical gap anisotropy of $470 \pm 15$ meV. This is the highest observed value for any material in the near-infrared-to-visible spectral region to the best of our knowledge. Energetically above, we identify multiple strongly polarized excitons spanning 1.25 eV to 3.1 eV selectively aligned along the two orthogonal axes. A resonance $X^-$, located 24 meV below the $X_0^b$ progressively transfers oscillator strength to $X_b^0$, a behavior consistent with a coupled trion (Fermi-polarion)/exciton pair. Our experiments also provide polarization-resolved broadband dielectric functions of CrSBr. These results establish CrSBr as a strongly polarization-selective excitonic system and highlight its potential for polarization-selective optoelectronics enabled with its large optical-gap anisotropy.

Two-dimensional (2D) van der Waals crystals combining strong in-plane anisotropy and magnetism have opened a new paradigm in the 2D research[1–7]. Crystals of the transition-metal chalcogen halides family such as CrSBr, CrOCl, CrOBr and FeOCl, and from the family $MAX_n$ ($M$ = Mn, Fe, Ni, Cr, Co; $A$ = P, Si, Ge, and $X$ = S, Se, Te, $n$ = 3,4) such as $CrPS_4$[8], $NiPS_3$[9], and $CrTe_2$[10] are emerging candidates with potential applications in polarized photodetectors[11,12], optical computing[13], anisotropic light-emitting diodes[14], magnon waveguides/transistors[15], and multi-bit read-only memories owing to magnon-transport anisotropy[8,16–19] and twist engineering[20].

Recently, CrSBr has attracted significant attention owing to its air stability and relatively high Néel temperature ($T_N$~132K)[1,21–24]. In its bulk form, CrSBr crystallizes in an orthorhombic structure belonging to the $Pmmn$ space group (No. 59)[1,21,22]. As illustrated in Fig. 1(a), a monolayer (1L) consists of corrugated Cr–S chains along the **b**-axis, sandwiched between top and bottom Br atoms. This imparts a pronounced anisotropic character to the crystal, effectively giving rise to a quasi-1D behavior[25–31]. As a result, monolayer and multilayer CrSBr exhibit strongly anisotropic optical and magnetic behavior[1,31]. Unlike most magnetic materials, CrSBr is a direct band gap magnetic semiconductor with minimum gap at Γ point of the Brillouin zone irrespective of the layer thickness[27]. Owing to the strong in-plane anisotropy, the lowest bright exciton in CrSBr is anisotropic as well[25–30,32]. Interestingly, CrSBr is reported to host two types of excitons with different spatial extent[33–35]. Overall, CrSBr is emerging as a highly promising platform for optical logic architectures and photonic computation, where the polarization state of light can be directly harnessed to encode binary information[36,37].

Within the layer plane, the exchange within each individual layer is ferromagnetic with magnetic easy axis along **b**[38–41]. The out-of-plane antiferromagnetic interlayer coupling leads to an A-type antiferromagnet below $T_N$[22,39,42]. Optical and magneto-optical spectroscopies have revealed an extremely rich physics in this material system such as excitons of large binding energy[43], trions/Fermi polarons[24,44], surface and bulk excitons[30,45], self-hybridized polaritons[46], spin-phonon coupling[47], and exciton-magnon coupling[19], and twist-tunable interlayer interactions[20].

Despite these recent developments, the polarized optical response of CrSBr is not well understood. For instance, the optical gap anisotropy, $\Delta E_{opt} = E_{onset}^a - E_{onset}^b$ where $E_{onset}^{a,b}$ are the lowest resolved bright excitonic absorption onsets for $E||\mathbf{a}$ or $E||\mathbf{b}$, is not known experimentally. Additionally, most prior experimental work is focused on the two lowest energy exciton resonances along the **b**-axis. However, the lowest bright optical onset along **a**-direction, although predicted theoretically[27], could not be detected using reflectance-contrast spectroscopy[25,30]. One experimental report has resolved an **a**-polarized exciton, although not the lowest-energy resonance for this polarization[29]. Furthermore, systematic polarization-resolved mapping of the excitonic landscape above 2 eV along both in-plane axes is still lacking. In addition to this, there is an ongoing debate on the identification of multiple optical resonances in negatively doped CrSBr layers around the lowest bright exciton. One work reports on the observation of a trion/Fermi polaron in doped trilayer CrSBr about 19 meV below the neutral exciton[24], while two other studies find a surface exciton state at a similar energy[30,48]. Differences also remain among the reported broadband dielectric functions of CrSBr[15,49]. Polarization-resolved reflectance, transmittance, and absorption measurements as a function of temperature together with $GW$-Bethe-Salpeter equation ($GW$-BSE) level *ab initio* calculations can help address these open questions.

In the present work, we perform polarization-resolved micro-absorption ($Abs$) on monolayer (1L) and 15 nm bulk-like

CrSBr in the near-infrared-to-visible spectral range spanning 1.25 eV to 3.1 eV. Alongside, we perform *GW*-BSE based *ab initio* calculations in this energy range. We resolve multiple closely-spaced excitonic transitions in the full energy region, each exhibiting strong in-plane polarization along distinct crystallographic directions. Temperature-dependent transmittance measurements reveal a transfer of oscillator strength between the two lowest energy neighboring resonances along the **b**-axis, consistent with a trion-exciton pair.

We perform our experiments on an hBN-encapsulated monolayer CrSBr and a 15 nm thick CrSBr crystal on **c**-cut 425 μm thick sapphire substrate for temperature between 9 K and 205 K. Absorptance ($Abs = 1 - Ref - Tr$) is measured by simultaneous measurement of reflectance ($Ref$) and transmittance ($Tr$) on the sample under the same conditions. The details of the experimental setup are provided in the supporting information. Thickness of the crystals was measured using atomic force microscope imaging. Monolayer thickness was also confirmed using photoluminescence spectroscopy[25,30] (see Fig. 1(b) and section SII of supporting information).

Figure 1(c) shows orthogonally polarized absorption spectra along the crystallographic **a**- and **b**-axes for 15 nm bulk-like and monolayer CrSBr at a temperature $T = 70$ K. The spectra exhibit multiple polarization-selective excitonic resonances along the two crystallographic axes. A direct comparison of the two crystallographic directions reveals a far richer *bright* excitonic landscape along the **b**-axis when compared with the **a**-axis (Fig. 1(c)). We notice that absorption reduces considerably with reducing thickness from bulk to monolayer. For instance, after integrating the absorptance over the measured spectral window from 1.25 to 3.1 eV and normalizing by the width of this window, i.e. taking the spectral average, 15 nm thick bulk-like crystal absorbs nearly 12 % and 18 % of the light polarized along the **a** and **b**-axes respectively. The corresponding numbers for the monolayer are about 1.5 % and 2 %, respectively. This is in agreement with a recent work where integrated absorption is predicted using generalized transfer matrix model, and an anomalous enhancement of absorption is found around this thickness range[50]. In comparison, for layered semiconductors of $MX_2$ ($M$ = Mo, W; $X$ = S, Se, Te) family, typical broadband normalized integrated absorption for monolayers has been found to be around 7.5 % for $WS_2$ and $MoSe_2$, while for around 15 nm, it is similar to CrSBr ($15 - 20$ %)[50].

To model the absorption spectra, we use normalized Lorentzians to represent all exciton features in the spectra. Furthermore, beyond the single-particle band gap, a step-like Sommerfeld-enhanced absorption background is present similar to other 2D semiconductors such as GaAs quantum wells[51,52], and $WS_2$[53]. Therefore, our model is represented by the following function overall:

$$I(E) = \frac{1}{2\pi}\sum_{i=1}^{m}\frac{A_i\Gamma_i}{(E-E_i)^2+\left(\frac{\Gamma_i}{2}\right)^2} + \sum_{j=1}^{n} S_j^{2D}(E) \qquad (1)$$

where $E_i$, $A_i$ and $\Gamma_i$ are the resonance energy, oscillator strength parameter and line width of the $i^{th}$ resonance. $S_j^{2D}$ represents the Sommerfeld-enhanced step-like function for the continuum corresponding to the $j^{th}$ exciton resonance. The fitted curves are shown in Figs. 1(c) and 1(d) as solid lines. A detailed example of the line shape fitting for both bulk-like and monolayer crystals along **a** and **b**-axes is given in Fig. S4 of the supporting information.

The lowest energy prominent **b**-polarized feature for both thicknesses is the well documented $X_0^b$ exciton[25] [see Fig. 1(c) and 1(d)]. It appears at $E_{X0}^b = 1.360$ eV (1.338 eV) for bulk (monolayer) at $T = 70$ K, and has been studied using

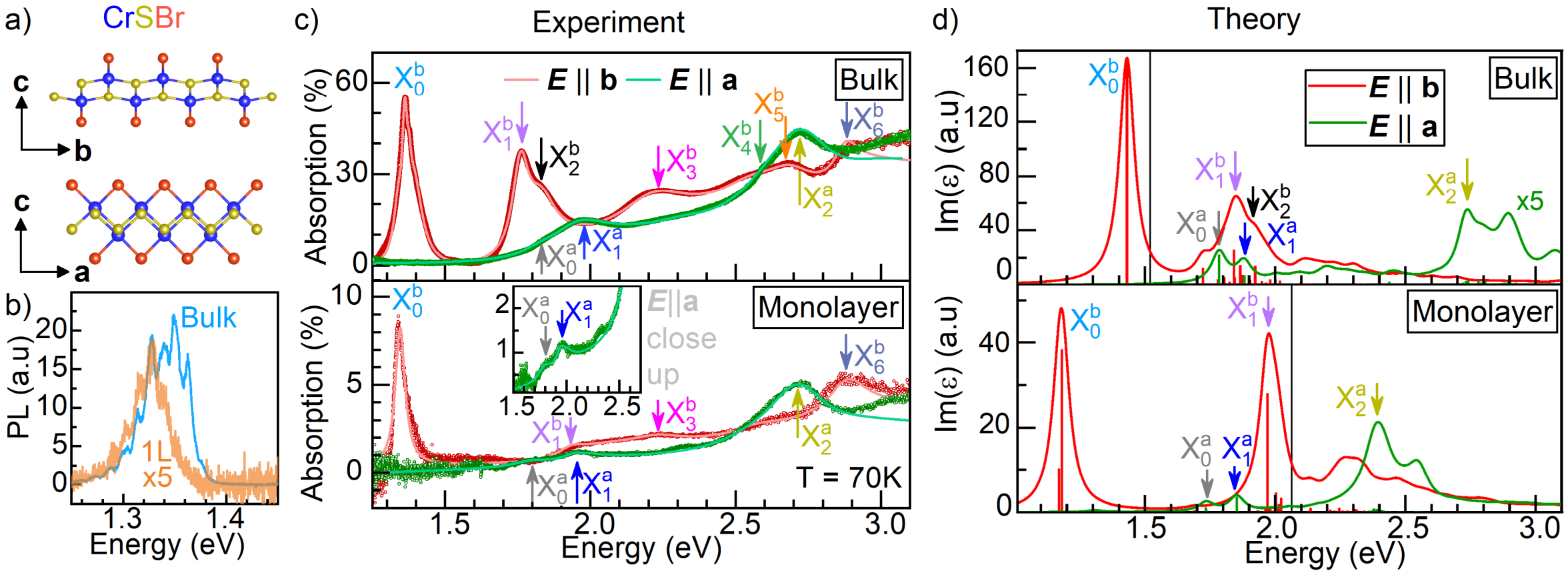

Figure 1. (a) CrSBr crystal structure in the **bc** and **ac** planes. Coordination of Cr (blue), S (yellow), and Br (brown) atoms is displayed. (b) Unpolarized PL spectra of bulk-like and monolayer CrSBr for 532 nm excitation, 1 mW power focused to 1.2 μm diameter spot. A red shift of the monolayer's PL compared to bulk is evident as found in literature[25,30]. (c) Red and green circles represent measured absorption ($1 - Ref - Tr$) spectra for detection polarization parallel to the **b**- and **a**-axes of bulk-like(top panel) and monolayer (bottom panel) CrSBr, while the corresponding red and green solid curves depict the modelled spectra. Inset in the monolayer case (bottom panel) is the zoomed-in view for $E||$**a**. $X_0^b$ and $X_0^a$ represent the lowest exciton absorption onsets for the two directions with $\Delta E_{opt}$~470 meV. These measurements are performed at $T = 70\ K$ temperature. (d) Theoretical optical absorption spectra using *GW*-BSE ab initio calculations. Red and green curves represent the calculated absorption spectra for $E||$**b** and $E||$**a**, respectively for both bulk (top panel) and monolayer (bottom panel) crystals.

Table 1. Experimentally reported polarization-dependent optical onsets and corresponding optical-gap anisotropy in anisotropic layered and conventional semiconductors.

| | Material | $E_1$ (eV) | $E_2$ (eV) | $\Delta E = E_2 - E_1$ (meV) |
|---|---|---|---|---|
| | *Layered Semiconductors* | | | |
| 1. | CrSBr (This work) | 1.36 (bulk) 1.338 (1L) | 1.83 (bulk) 1.81 (1L) | 470 |
| 2. | $[ReS_2]$[65] | 1.5 to 1.7 eV depending on thickness | | 45 – 80 |
| 3. | $[ReSe_2]$[13] | 1.36 to 1.55 eV depending on thickness | | 20 – 40 |
| 4. | $[ZrS_3]$[66] | 2.055 | 2.085 | 30 |
| 5. | SnSe[67] | 1.04 | 1.10 | 60 |
| 6. | SiAs[68] | 1.48 | 1.57 | 90 |
| 7. | GeS[69] | 1.70 | 1.80 | 100 |
| 8. | SnS[70] | 1.28 | 1.48 | 200 |
| | *Conventional Semiconductors* | | | |
| 9. | $[\beta\text{-}Ga_2O_3]$[71] | 4.5 | 4.6 | 100 |
| 10. | $[Cu_2ZnSiS_4]$[72] | 3.345 | 3.432 | 87 |
| 11. | $[Cu_2ZnSiSe_4]$[72] | 2.348 | 2.406 | 58 |
| 12. | a-InN[73] | 0.73 | 0.75 | 20 |
| 13. | $[Al_xGa_{1-x}N$ $x \approx 0.13\text{-}0.8]$[74] | 3.5 to 5.3 eV depending on x | | 15 – 170 depending on x |
| 14. | m-AlN[75] | 5.94(‖**c**) | 6.19 (⊥**c**) | 220 |
| 15. | a-GaN[76,77] | 3.432 | 3.441 | 9 |
| 16. | m-GaN[78] | 3.428 | 3.468 | 40 |
| 17. | Semi polar GaN[79] | 3.401 | 3.428 | 27 |
| 18. | $a\text{-}Al_{0.81}In_{0.19}N$[80] | 4.20 | 4.34 | 140 |
| 19. | a-ZnO[81,82] | 3.420 | 3.449 | 30 |
| 20. | CdS[83] | 2.54 | 2.56 | 20 |
| 21. | [Anatase $TiO_2$][84] | 3.79 | 4.13 | 340 |

*Footnote: $E_1$ and $E_2$ denote the lower- and higher-energy experimentally resolved optical onsets for orthogonal linear polarizations, respectively; $\Delta E = |E_2 - E_1|$. For CrSBr, $E_1 = E^b_{X0}$ and $E_2 = E^a_{X0}$* polarized photoluminescence and reflectance previously[25]. This resonance therefore defines the **b**-polarized optical gap in the terminology used here. This transition is assigned as the lowest bright excitonic absorption at the Γ point of the Brillouin zone, consistent with earlier reports[17,25,27,30,34,35]. Along the **a**-axis, the lowest energy transition $X^a_0$ is found at $E^a_{X0} = 1.83$ eV (1.81 eV) for bulk (monolayer). This resonance is visible only in transmittance or absorption spectra, and is absent in the reflectance spectrum (see Fig. S5 of the supporting information), consistent with a recent work[30]. The absence of this weak feature in reflectance underscores the advantage of simultaneously measured R and T for determining absorptance. To our knowledge, this resonance has not previously been reported. A recent study instead measured the next higher-energy exciton $X^a_1$ (at 1.97 eV, discussed in the next para) using photoluminescence excitation spectroscopy[29]. Therefore, for both monolayer and bulk-like CrSBr, we experimentally find an optical gap anisotropy of $\Delta E^{a-b}_{Exp} = E^a_{X0} - E^b_{X0} = 470 \pm 15$ meV at $T = 70$ K. To the best of our knowledge, this is the largest experimentally reported optical-gap anisotropy for any material in the near-infrared-to-visible spectral region, as summarized in Table 1. This highlights CrSBr as a promising platform for spectrally separated, ultrathin, polarization-selective optical states.

Apart from the lowest bright resonances $X^a_0$ and $X^b_0$ which define the lowest bright optical onsets along the two axes in the bulk-like and monolayer crystals, we notice multiple **a** and **b** polarized excitons. In bulk-like CrSBr (Fig. 1(c)) with polarization along **b**-axis, six prominent features appear around $E^b_{X1} = 1.76$ eV, $E^b_{X2} = 1.830$ eV, $E^b_{X3} = 2.23$ eV, $E^b_{X4} = 2.58$ eV, $E^b_{X5} = 2.68$ eV, and $E^b_{X6} = 2.88$ eV. Out of these, $X^b_1$ and $X^b_2$ have been reported recently[33,35], where their behavior under magnetic fields has been studied. One of these reports suggests that $X^b_0$ and $X^b_1$ exhibit a slightly different extension in real space [33]. On the other hand, for monolayer, only $X^b_1$ (1.936 eV), $X^b_3$ (2.23 eV) and $X^b_6$ (2.88 eV) are discernible, around similar energies as that of bulk. Now we turn our attention to excitons polarized along the **a**-axis for bulk and monolayer, respectively. In both cases, we detect two **a**-polarized exciton resonances $E^{a}_{X1} = 1.97$ eV and $E^{a}_{X2} = 2.72$ eV. The resonance $X^{a}_{1}$ has been reported recently, consistent with our observations[29].

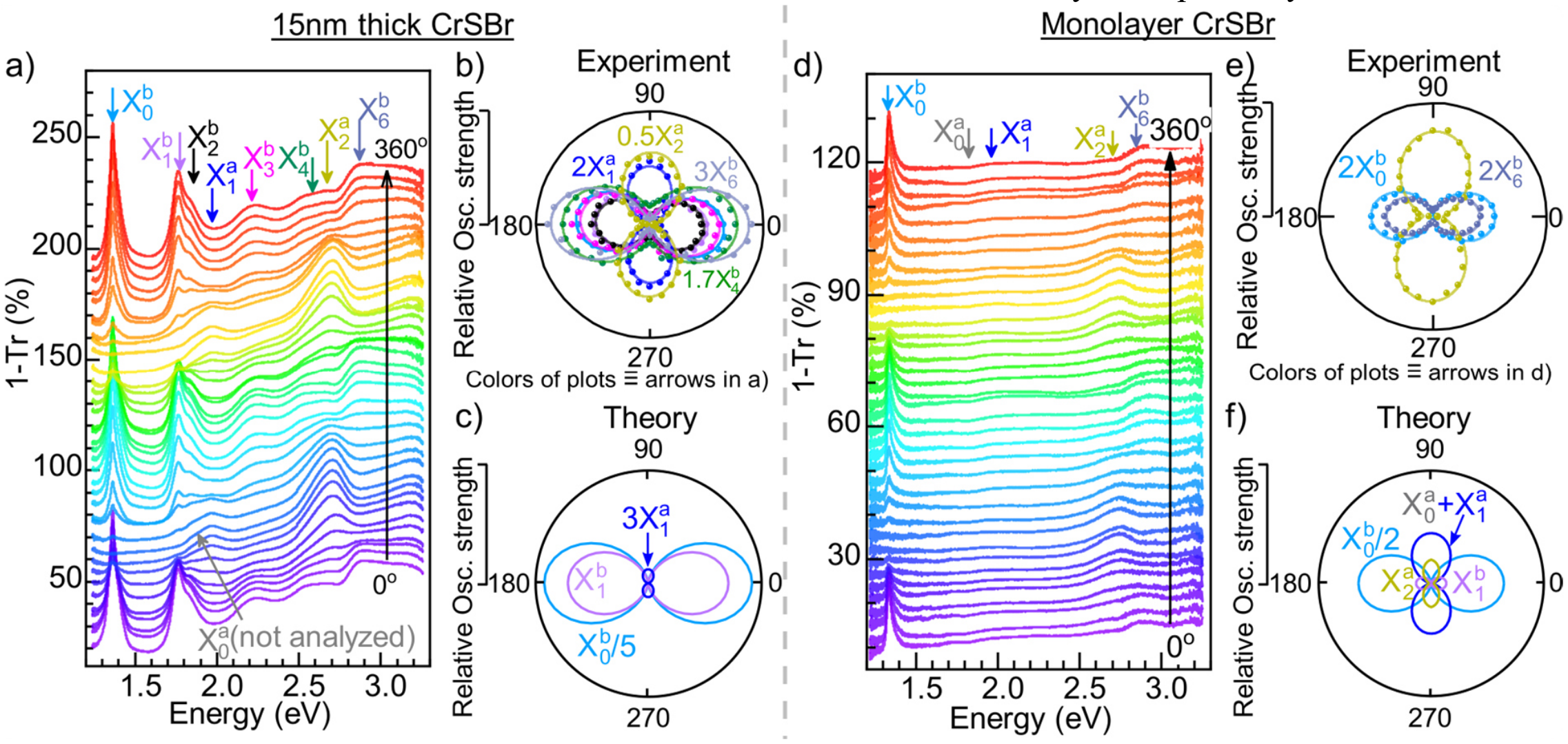

Figure 2. (a) Experimentally measured polarization resolved $1 - Tr$ spectra for bulk with 0° ≡ **b**-axis and 90° ≡ **a**-axis. The spectra are vertically shifted successively with respect to the 0° spectrum (≡**b**-axis) for clarity. (b) Experimentally determined relative spectral weights of excitonic resonances for bulk-like CrSBr. (c) Calculated excitonic oscillator strengths for a few excitonic transitions using *GW*-BSE based *ab initio* theory. (d) – (f) are similar to (a) – (c), respectively, but for monolayer CrSBr.

Our theoretical investigations (Fig. 1(d)) show a similar trend with a large series of excitons with different polarizations. The important details of the calculations are provided in Section SIV of the supporting information. We discuss the case of the bulk CrSBr first. The first **b**-polarized exciton $X^{b}_{0}$ in bulk is found at $E^{b}_{X0} = 1.43$ eV, while the first resonance with strong weight in **a** ($X^{a}_{0}$) is located at $E^{a}_{X0} = 1.78$ eV. While $X^{b}_{0}$ results from transitions close to $\Gamma$, $X^{a}_{0}$ is localized close to the $X$ point in reciprocal space. The corresponding calculated splitting for bulk CrSBr i.e. $\Delta E^{a-b}_{Th} = E^{a}_{X0} - E^{b}_{X0} = 350$ meV is somewhat smaller than the experimental value of 470 meV. The $X^{b}_{0}$ and $X^{a}_{0}$ excitons are individually converged to better than 10 meV because they lie below the corresponding direct single-particle gaps at $\Gamma$ (1.52 eV) and close to $X$ (about 2 eV) points, respectively. Excitonic resonances above these respective continuum thresholds are less well converged. The observed agreement should therefore be regarded as reasonable. Note that the lowest single-particle interband transition at $\Gamma$, $VB \rightarrow CB1$, is dipole forbidden in our calculated band ordering. This is not observed experimentally. However, the lowest bright exciton ($X^{b}_{0}$) is a transition from VB to the $2^{nd}$ conduction band (CB2) which is higher in energy than CB1 by $E^{CB2} - E^{CB1} = 150$ meV. The corresponding band-to-band bright transition $\mathrm{VB} - \mathrm{CB2}$ takes place at 1.67 eV. The separation between $X^{b}_{0}$ and its corresponding dipole-allowed VB–CB2 continuum is approximately 240 meV, which we identify as the binding energy of this bright excitonic channel. For energies above 2 eV, we find many excitations in agreement with experiment. However, these transitions are more challenging to analyze due to their nature (extended in real space, described by a series of vertical lines in the spectrum) and a one-to-one comparison with experiment is challenging.

For monolayer CrSBr, our calculations reveal an electronic band gap of about 2.06 eV, while the corresponding energy difference of the conduction bands is $E^{CB2} - E^{CB1} \sim 35$ meV. A stronger exciton binding energy of about 0.9 eV results in $X^{b}_{0}$ at about 1.2 eV polarized along **b**. The first exciton polarized along **a** $X^{a}_{0}$ is at about 1.75 eV, leading to an anisotropy of $\Delta E^{a-b}_{Th} = E^{a}_{X0} - E^{b}_{X0} = 550$ meV for the monolayer. At higher energies, the calculated spectrum reproduces the predominantly **a**- or **b**-polarized character of the experimental resonances, although the individual peak energies show deviations because of the more challenging convergence in this above-gap regime. Note that in comparison to the sapphire substrate in experiment, no substrate is present in our calculations. This may explain some discrepancy of the theoretically calculated exciton energies with respect to the experimental measurements[54].

To obtain the complete polarization characteristics of the excitons, we perform polarization-resolved transmittance ($Tr$) spectroscopy on our samples from 0°|| **b**-axis to 360° in steps of 10°. Here, 90° polarization corresponds to the **a**-direction. We plot $1 - Tr$ spectra in Fig. 2(a) and 2(d) for bulk-like and monolayer crystals, respectively. It is noteworthy that measuring true absorption i.e. $Abs = 1 - Ref - Tr$ is only possible for the two orthogonal polarizations such as in Fig. 1(c). However, for other angles, polarization of reflected light is spoiled because of a beam splitter as shown in Fig. S1 of the supporting information, due to which only $1 - Tr$ is measured in Fig. 2. The $1 - Tr$ spectra approximately correspond to the absorption spectra since reflectance contributions to the excitonic $1 - Tr$ line shapes are relatively much smaller. Reflectance indeed introduces a featureless background which is subtracted during analysis (see Fig. S5 of the supporting information for individual $Ref$ and $Tr$ spectra). From the spectral modelling in eq. 1, we extract relative spectral-weight parameters for all resonances in Figs. 2(b) and 2(e). Some transitions, namely $X^{a}_{0}$ ($X^{a}_{0}$ and $X^{a}_{1}$) for bulk (monolayer), however, could not be analyzed for polarization dependence. We clearly notice a preferential orientation of the analyzed exciton transitions along **a** or **b**-axes, in almost perfect dipole patterns. These oscillator strength plots of the transitions are fitted using the following function[13]

$$A = A_x \cos^2(\theta - \theta_0) + A_y \sin^2(\theta - \theta_0) \qquad (2)$$

where $A_x$ and $A_y$ are the **b** and **a**-polarized components of the oscillator strength respectively, and $\theta_0$ is the orientation of the dipole with respect to **b**-axis. As expected from visual inspection of the plots, the obtained values of $\theta_0$ are either $0° \pm 4°$ or $90° \pm 4°$. We note that for the bulk-like crystal, $X^{b}_{2}$ and $X^{a}_{0}$ appear at nearly same energy, making it difficult to explicitly determine their individual polarization-dependent oscillator strength. Through our theoretical calculations (Figs. 2(c) and (f)), we plot the polarization of some excitons as labeled in Figs. 1(d). The extracted experimental spectral weights in Figs. 2(b) and (e) exhibit the same strong axis selectivity as the theoretically calculated excitonic oscillator strengths. As discussed for the energetic positions, the relative oscillator strength follows similar trends as in experiment while a one-to-one comparison of each single peak remains challenging.

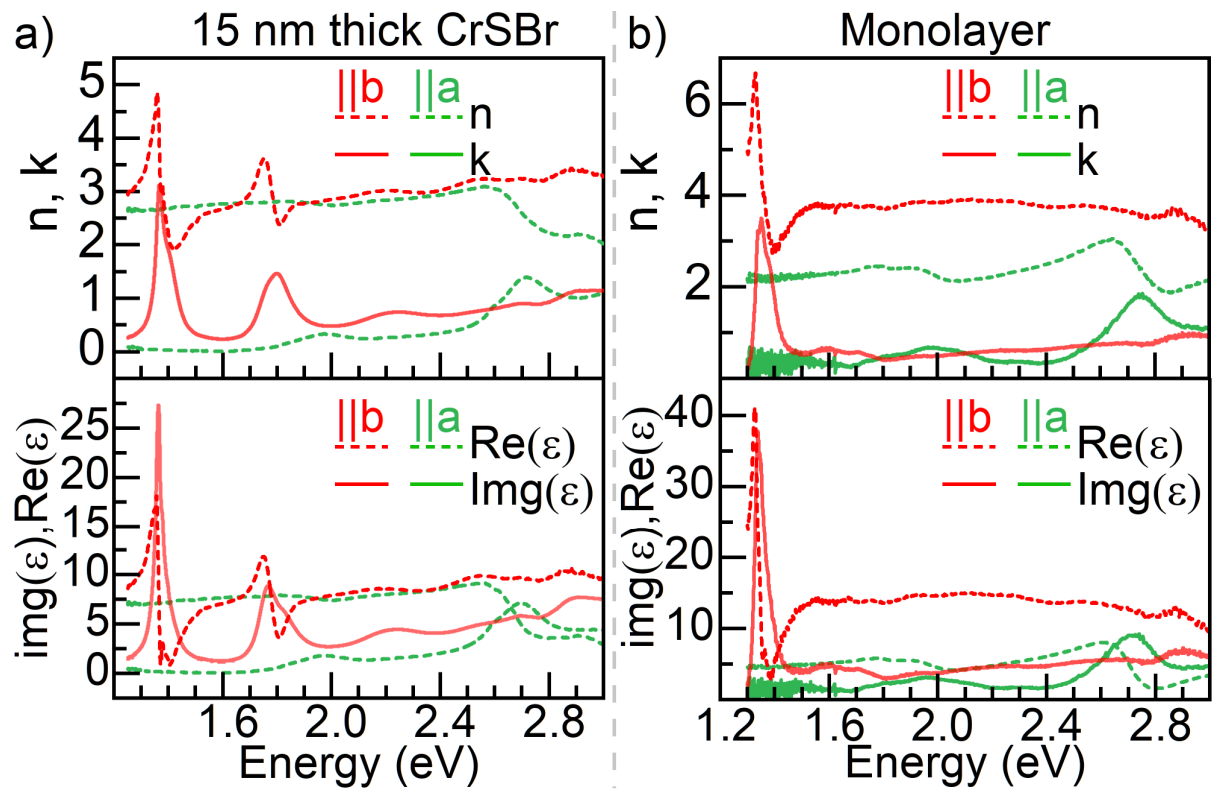

Figure 3. Extracted polarized optical constants of CrSBr. (a) Complex refractive index (n, k) (upper panels) and complex dielectric function (Re(ε), Im(ε)) (lower panels) for the 15 nm thick bulk-like CrSBr extracted from the experimental data using the generalized transfer matrix method (GTMM). (b) Corresponding complex refractive index and dielectric function for monolayer CrSBr extracted using GTMM.

The analysis of our experimental results also provides information on the dielectric functions of CrSBr, which is a matter of debate currently[15,49]. Figure 3 presents the polarization-dependent (**a**- and **b**-axes) optical constants-the complex refractive index ($\tilde{n} = n + ik$) and complex dielectric function ($\epsilon = \tilde{n}^2 = \epsilon_1 + i\epsilon_2$) for 15 nm thick and monolayer CrSBr. The values of $n$ and $k$ are extracted by simultaneously fitting experimental reflectance and transmittance spectra using a generalized transfer-matrix model (GTMM) of the air/CrSBr/sapphire (air/7nm hBN/CrSBr/62nm hBN/Sapphire) system for bulk (monolayer), as explained in Ref. [50]. We used the incoherent substrate approximation to extract the refractive indices[55]. To improve the robustness of the inversion, the fits were constrained using the Kramers-Kronig formalism[56]. We note that the refractive indices derived in our work are significantly different from those reported for a $> 100$ nm thick CrSBr recently[15], where a refractive indices $n$ and $k$ reaching approximately 25 near the exciton resonance are reported. There could be many reasons for this discrepancy: 1) we calculate dielectric functions from combined reflectance and transmittance measurements, while only differential reflectance (DR) is performed in[15] which could lead to large uncertainties (low convergence of dielectric functions), 2) our CrSBr crystals are kept on sapphire substrate while in [15], the samples are either placed on 90 nm $SiO_2$/Si, or are embedded within a distributed Bragg-reflector based cavity where interference effects dominate[57], 3) different thicknesses of CrSBr used in both works might affect the dielectric functions. However, such strong dependence of dielectric functions on thickness is not usually expected in layered semiconductors[50]. In another recent work based on spectroscopic ellipsometry of 65 nm thick CrSBr[49], we also notice some qualitative differences, while the magnitude of refractive indices is similar to the present work. For instance,[49] reports some extra low-in-energy **a**-polarized resonance-like features in the dielectric function. Our analysis therefore provides complementary low-temperature polarization-resolved optical constants for thin and monolayer CrSBr, obtained from simultaneous reflectance and transmittance measurements. The remaining differences between reported dielectric functions may ultimately be resolved through systematic micro-ellipsometry and absorption measurements as a function of layer thickness.

We now turn our attention to the fine structure we observe around the absorption line of the lowest bright resonance $X_0^b$. In Fig. 4(a), we show zoomed-in view of the $X_0^b$ spectral feature ($1 - Tr$) measured at $T = 9$ K. At the low-energy side of $X_0^b$ ($E_{X0}^b = 1.370$ eV), a prominent feature $X^-$ is observed at $E_{X^-} = 1.346$ eV. As we explain in the next paragraph, we tentatively assign this satellite to a trion/Fermi-polaron with an energetic separation $E_{X0}^b - E_{X^-} = 24$ meV[24,44]. Within a trion interpretation, this separation corresponds to a binding energy of 24 meV. At the high-energy side of $X_0^b$, multiple spectral lines are detected which we associate with phonon replicas of $X^-$ and $X_0^b$. In the following, we provide evidence for our associations.

A well-known optical fingerprint of a trion-exciton pair in the absorption spectrum of a doped semiconductor involves the behavior of the pair as a function of temperature[53,58,59]. When the temperature of the system is raised, the trion loses its oscillator strength while the exciton gains. This oscillator-strength transfer as a function of temperature is explained in Fig. 4(c)[53,58,60]. At low temperatures, the excess carriers in a doped semiconductor (assume $n$ doping) occupy states close to the bottom of the conduction band (at $\Gamma$ point for CrSBr) within the light cone. This favors the optical creation of trions which require excess carriers, resulting in a significant oscillator strength. However, since the phase space around the $\Gamma$ point in conduction band is Pauli blocked, the creation of neutral excitons is not favored, leading to a relatively lower oscillator strength. As the temperature increases, the carriers in the conduction band are thermally redistributed to higher $k$ points according to Fermi-Dirac distribution, outside the light cone. Due to this, the trions lose oscillator strength, while the excitons gain. In our case of bulk-like CrSBr, we notice this behavior for the $X^-/X_0^b$ pair in Fig. 4(b) and (d) for temperature between 9 K and 135 K beyond which the trion is not detectable. A surface-exciton contribution cannot be excluded[30,48]; nevertheless, the observed temperature-dependent spectral-weight transfer is consistent with a charged-exciton/Fermi-polaron assignment. This closely resembles the behavior reported for doped few-layer CrSBr[24] which discusses charged excitons red shifted by 19 meV from $X_0^b$ in a doped trilayer CrSBr. This work also reports another exciton $X_B'$ red shifted by 5 meV from $X_0^b$ (denoted by $X_B$ in[24]) in trilayer CrSBr. The authors conclude that $X_B'$ is localized in the middle layer of the trilayer, while $X_0^b$ is in the outer layers (similar to a surface exciton). Such doublets are common in semiconducting transition-metal dichalcogenide multilayers as well such as in $MoS_2$[61]. We believe that this aspect should be explored further using gated CrSBr in combination with temperature and magnetic field-dependent absorption measurements, as a function of layer thickness, for settling this debate.

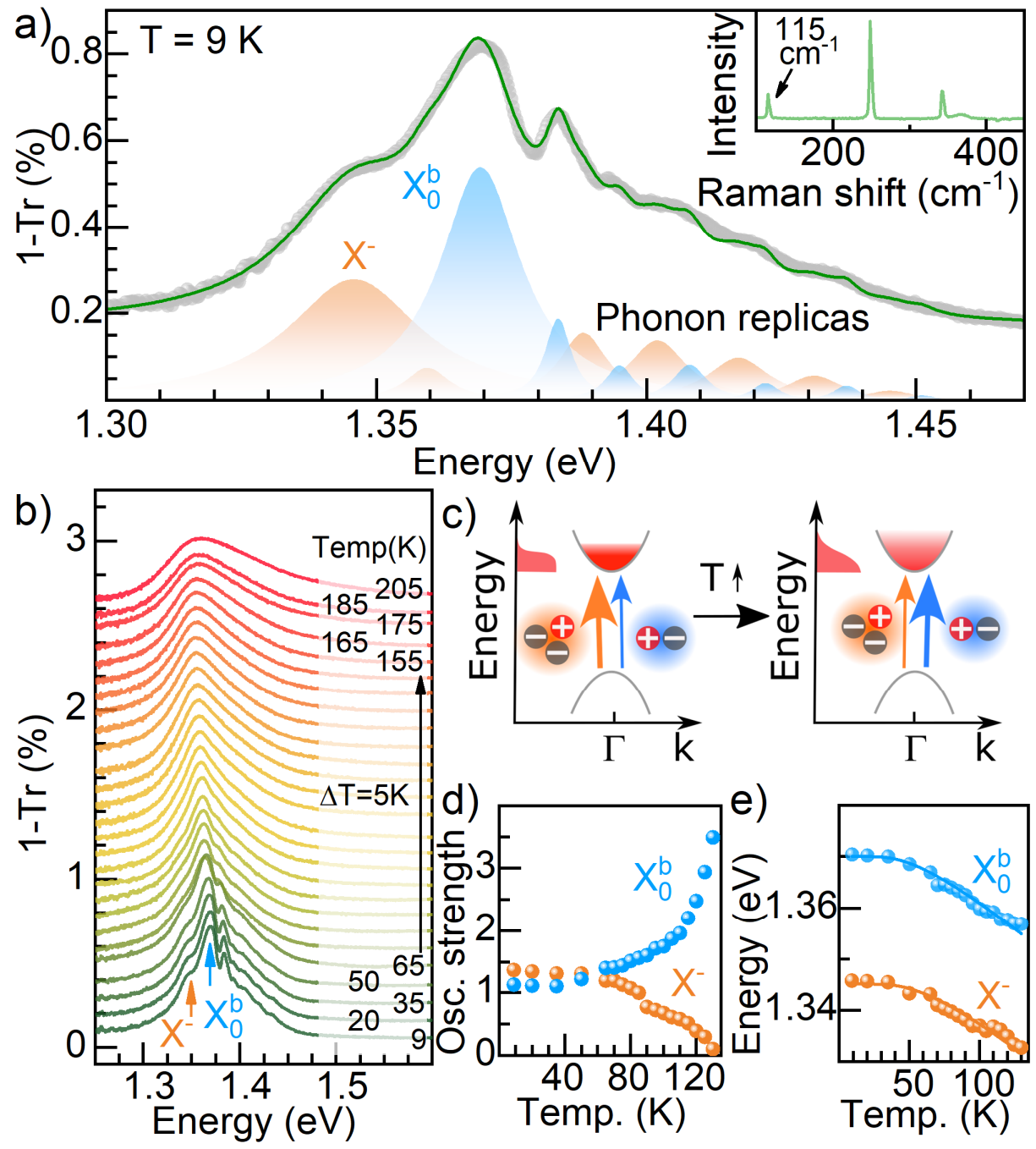

Figure 4. (a) Experimentally measured $1 - Tr$ spectrum showing equally spaced phonon replicas of $X^-$ and $X_0^b$. The inset shows the $A_g^1$ Raman peak (115 cm$^{-1}$) responsible for phonon replicas[62]. (b) Temperature dependent $1 - Tr$ spectra in the 9 K to 205 K temperature range. Spectra are vertically shifted for clarity. (c) Schematic drawing visualizing the transfer of optical weight between neutral excitons (blue vertical arrow) and trions (orange vertical arrow) with increasing temperature due to the thermal redistribution of carriers (red) in momentum space. (d) Extracted relative oscillator strength, and (e) transition energies of $X^-$ and $X_0^b$ from 9 K - 135 K beyond which $X^-$ cannot be reliably resolved.

The appearance of $A_g^1$ phonon replicas of $X_0^b$ with an energy separation of $115\ \mathrm{cm}^{-1} \cong 14\ \mathrm{meV}$ is not new[62]. However, we are not able to fit absorption spectrum in Fig. 4(a) using only one family of phonon replicas of $X_0^b$[62]. Instead, we empirically require another family corresponding to that of $X^-$ for obtaining a reasonable fit (solid line in Fig. 4(a)). The fit features corresponding to the two families of phonon replicas are shown as filled blue and orange peaks in Fig. 4(a). The strength of successive phonon replicas decreases with increasing replica order, consistent with previous observations in CrSBr[62]. To further confirm the coupling of the $A_g^1$ phonon with $X_0^b$ and $X^-$, we plot resonance energies of $X_0^b$ and $X^-$ as a function of temperature from $9\ \mathrm{K} - 135\ \mathrm{K}$ (see Fig. 4(e)). We fit these curves using the well-known O'Donnell's formula[63]

$$E_g(T) = E_0 - S\langle\hbar\omega\rangle\left[\coth\left(\frac{\langle\hbar\omega\rangle}{2k_BT}\right) - 1\right] \quad (3)$$

where $\langle\hbar\omega\rangle$ is the average phonon energy and $S$ is the coupling parameter. From our fits, we find $\langle\hbar\omega\rangle_{X_0^b} = 14 \pm 2\ \mathrm{meV}$ and $\langle\hbar\omega\rangle_{X^-} = 15 \pm 2\ \mathrm{meV}$ which is consistent with the $A_g^1$ energy.

In summary, polarization-resolved absorption spectroscopy reveals an optical-gap anisotropy of $470 \pm 15\ \mathrm{meV}$, defined by the splitting between the lowest bright **a**- and **b**-polarized excitonic absorption onsets, in both bulk-like and monolayer CrSBr. To the best of our knowledge, this is the largest value observed for a material in the near-infrared-to-visible spectral region. Supported with *GW*-BSE *ab initio* calculations, we identify many strongly polarized excitonic resonances in the energy range of 1.25 eV to 3.1 eV. Furthermore, we determine polarized complex refractive index/dielectric functions of both bulk and monolayer CrSBr. A low-energy satellite and associated phonon-replica structure near the lowest **b**-polarized exciton in bulk-like crystal show behavior consistent with a charged exciton/Fermi-polaron contribution. The giant splitting of the bright optical onset and the broadband axis-selective exciton landscape make CrSBr a compelling platform for polarization-selective and magneto-optical nanophotonics.

**Materials and Methods.** Details on experimental methods, sample preparation and characterization are provided in the supporting information.

**Data Availability Statement.** The raw data procured and analyzed in this work are available with the corresponding author on reasonable request.

**Supporting Information.** Supporting information is available free of charge at https://pubs.acs.org.

Detailed description of experimental setup; sample preparation and characterization using atomic force microscope, photoluminescence and Raman spectroscopy; examples of line shape modeling; individual reflectance/transmittance/absorption spectra on CrSBr crystals

We thank T. S. Mahesh, Sandip Ghosh, Aparna Deshpande, G. V. Pavan Kumar, Shouvik Datta, Seema Sharma and Sourabh Dube for support with equipment while building the experimental setups, and Arundhati Adhikari and Aparna Shinde for useful discussions. We acknowledge financial support from the following projects funded by the Government of India: NM-ICPS of the DST through the I-HUB Quantum Technology Foundation (Pune, India), DST Project No. CRG/2022/007008 of SERB, DST project No. ANRF/ARG/2025/002185/PS of ANRF, MoE-STARS project No. MoE-STARS/STARS-2/2023-0912, CEFIPRA CSRP Project No. 7104-2, VAIBHAV fellowship number INAE/DST-VF/2024/I/03, and DST National Quantum Mission project No. DST/QTC/NQM/QMD/2024/4 (G). Z.S. was supported by ERC-CZ program (project LL2101) from Ministry of Education Youth and Sports (MEYS), by project LUAUS25268 from Ministry of Education Youth and Sports (MEYS) and by the project Advanced Functional Nanorobots (reg. No. CZ.02.1.01/0.0/0.0/15-003/0000444 financed by the EFRR). T.D. acknowledges financial support from the Deutsche Forschungsgemeinschaft (DFG, German Research Foundation) through Project No. 426726249 (DE 2749/2-1 and DE 2749/2-2). The authors gratefully acknowledge the Gauss Centre for Supercomputing e.V. (www.gauss-centre.eu) for funding this project by providing computing time through the John von Neumann Institute for Computing (NIC) on the GCS Supercomputer JUWELS[64] at Jülich Supercomputing Centre (JSC).

**References.**

Supporting Information

# Extreme Polarization of the Optical Gap and High-Energy Exciton Landscape in CrSBr

Sayantan Patra[1], Sourabh Jain[1], Bhumika Chauhan[1], Marie-Christin Heißenbüttel[3], Abhisek Saidarsan[1], Ranjuna M. K.[1], Kseniia Mosina[2], Zdeněk Sofer[2], Michael Rohlfing[3], Thorsten Deilmann[3], Ashish Arora[1,*]

[1]*Department of Physics, Indian Institute of Science Education and Research, Dr. Homi Bhabha Road, 411008 Pune, India*
[2]*Department of Inorganic Chemistry, University of Chemistry and Technology, Prague 6, Technicka 5, 16628, Prague, Czech Republic*
[3] *Institute of Solid-State Theory, University of Münster, D-48149 Münster, Germany*
Email: ashish.arora@iiserpune.ac.in

## SI. Experimental set-up.

For polarization dependent reflectance and transmittance measurements, a Xenon lamp is used to cover the spectral range from 400 nm to 800 nm, while a tungsten-halogen lamp shines light for wavelengths between 800 nm and 1100 nm. For transmittance measurements, unpolarized light is focused onto the sample via an achromatic doublet ($f = 100$ mm). The transmitted light through the sample is collected by a 50x objective lens (N.A = 0.55). For reflectance, the same light source is directed through the 50x objective, which serves both to focus the incident and collect the back-reflected signal from the sample. In both configurations, the collected light passes through a linear analyzer to select the desired polarization. To achieve high spatial resolution, the output is focused on a pinhole of diameter 20 µm which selects 1 µm spot of the sample. After that the light is coupled into the spectrometer to acquire the spectrum. Reflectance is measured only for two orthogonal polarizations, whereas transmittance is measured between 0° to 360° in steps of 10°.

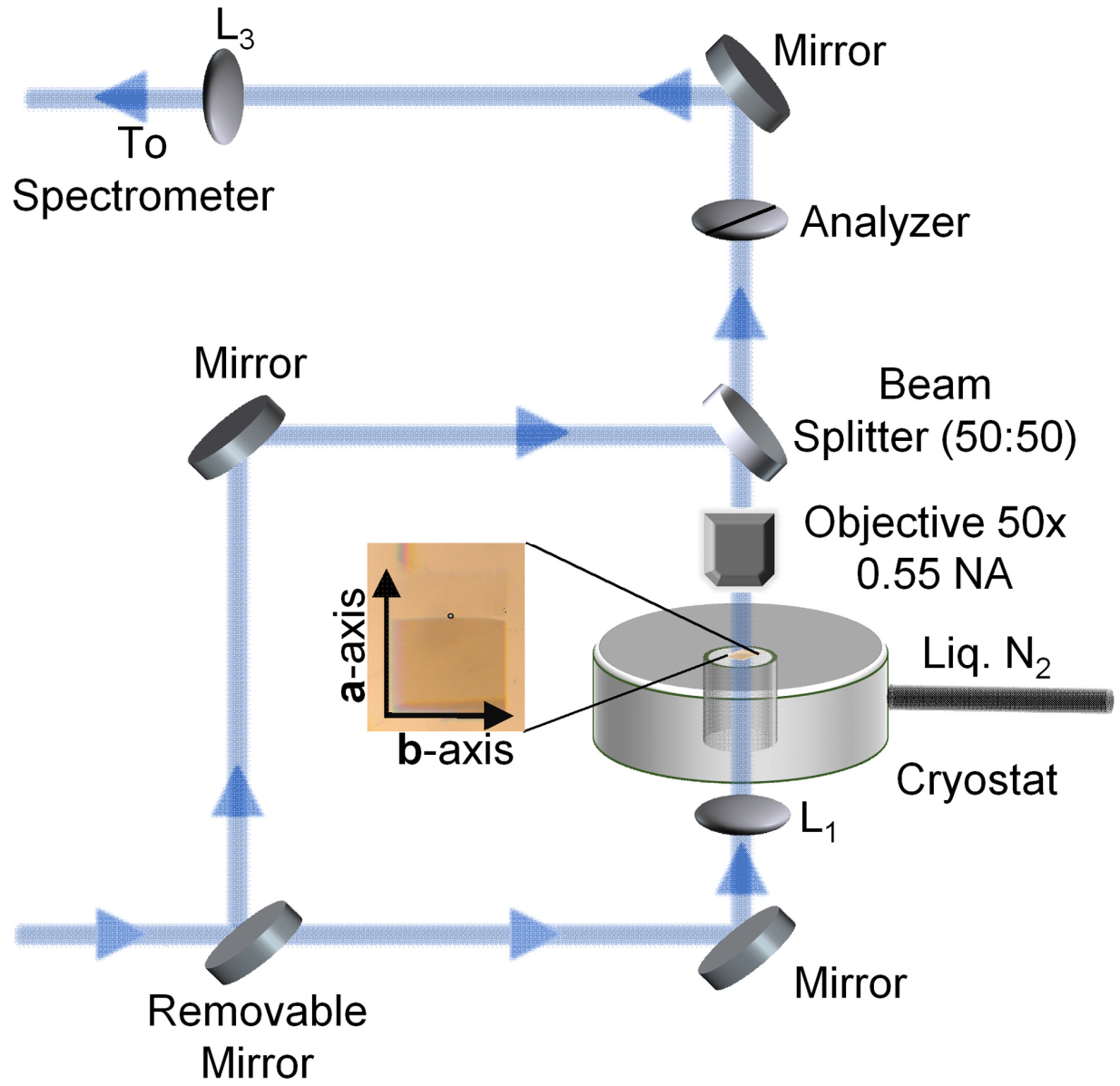

Figure S1. **Schematic of the optical spectroscopy setup.** The details are described in the accompanying text in this document.

## SII. Sample preparation and optical/AFM characterization.

Thin layers of CrSBr are mechanically exfoliated from a bulk crystal and transferred onto a 0.425 mm thick, polished c-cut sapphire substrate using a well-established dry-transfer method[1]. The monolayer crystal was encapsulated between thin layers of hBN for better optical response. We performed AFM measurements and cryogenic (T = 70K) photoluminescence (PL) spectroscopy to determine the flake thicknesses of CrSBr. For PL measurements we use a 532 nm laser for the excitation with 1 mW focused power using a 0.55 NA objective lens. Our measured optical response for monolayer CrSBr shows excellent agreement with previously reported results[2,3]. To further characterize the samples, we perform Raman spectroscopy on both monolayer and bulk CrSBr at 70 K using 633 nm excitation laser. Three prominent Raman-active phonon modes were observed at approximately $115\ \mathrm{cm^{-1}}$, $245\ \mathrm{cm^{-1}}$ and 346 $\mathrm{cm}^{-1}$ corresponding to the $A_g^1$, $A_g^2$ and $A_g^3$ out of plane vibrational mode respectively in excellent agreement with the previous studies[4–6].

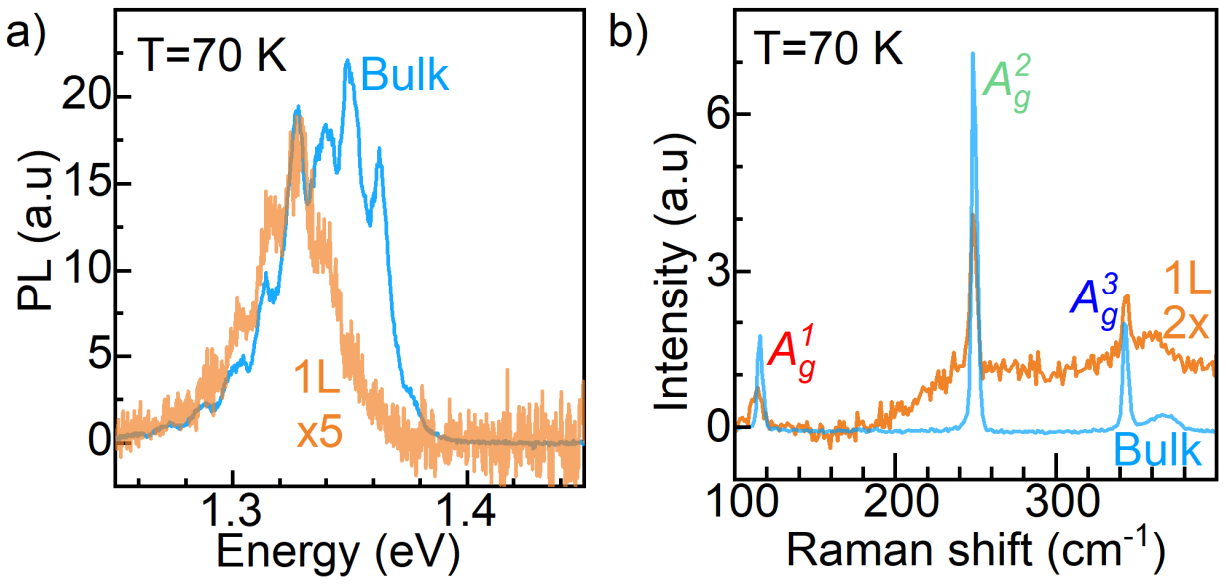

Figure S2. **Optical characterization of monolayer and 15 nm thick CrSBr.** a) Unpolarized low temperature (70K) photoluminescence spectra of monolayer (1L) and bulk-like (15nm thick) CrSBr for 532 nm excitation. b) Unpolarized low temperature (70 K) Raman spectra of monolayer (1L) and bulk-like (15nm thick) CrSBr for 633 nm excitation.

## SIII. Temperature-dependent Raman spectroscopy.

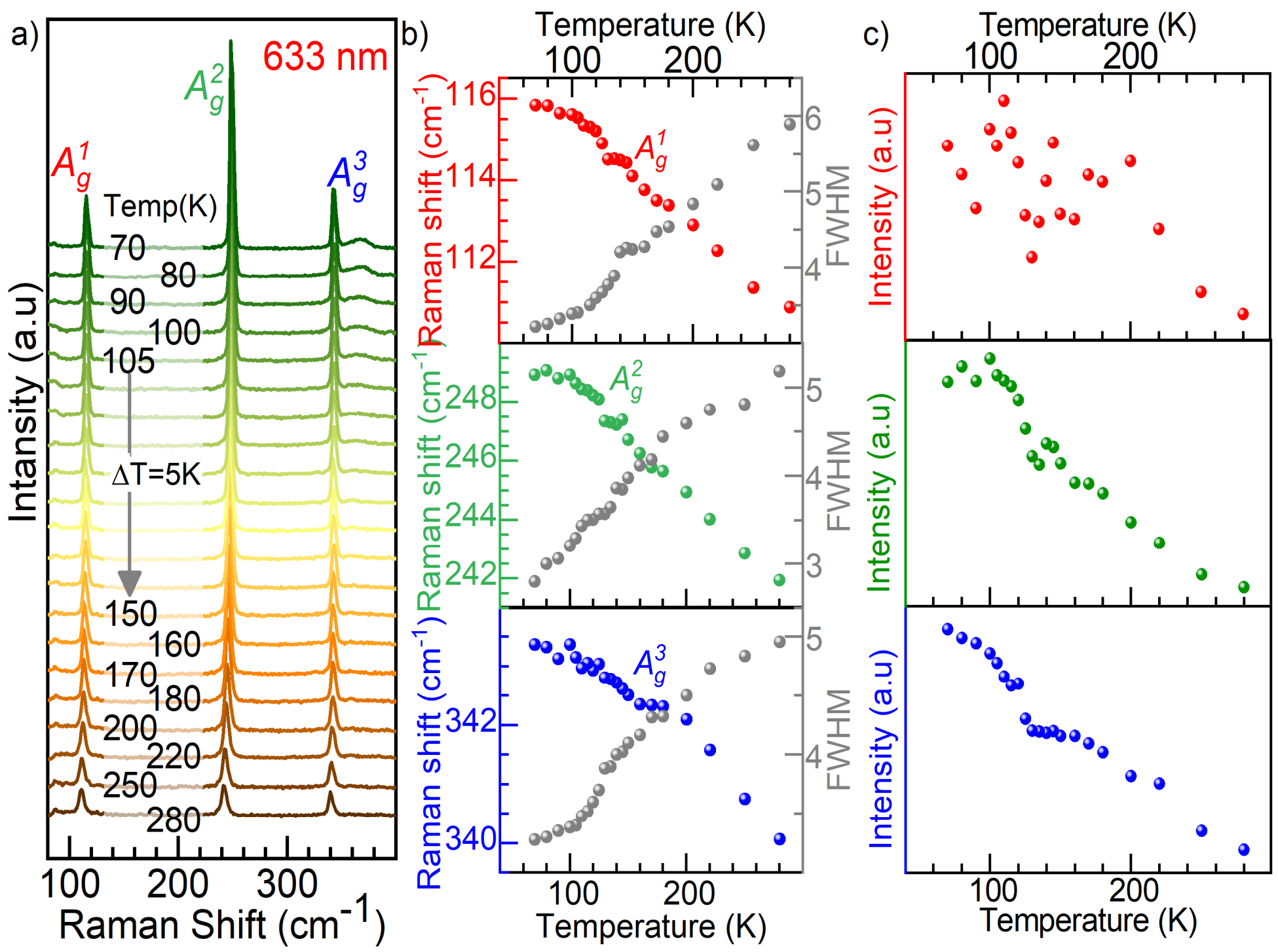

Figure S3. **Temperature-dependent Raman spectroscopy for 15 nm thick CrSBr under 633 nm excitation. (a)** Temperature-dependent Raman spectra measured from 70K to 280 K showing three characteristics out of plane modes $A_g^1$, $A_g^2$ and $A_g^3$. **(b)** Evolution of the peak frequency (Raman shift) and FWHM with temperature and **(c)** Corresponding integrated Raman intensity versus temperature for each respective mode.

The temperature-dependent Raman spectra of 15 nm thick bulk CrSBr are measured using 633 nm excitation laser over the temperature range 70 K to 280 K. We observe similar trends for all three Raman modes ($A_g^1$, $A_g^2$ and $A_g^3$): (i) All Raman modes show hardening (blue shift) from room temperature down to 70 K. For $A_g^3$ (346 $\mathrm{cm}^{-1}$) the phonon energy stays mostly unchanged in the temperature range $T_c = 160$ K to $T_N = 132$ K. It corresponds to the intermediate magnetic phase in which the interlayer ferromagnetism is established[5]. However we do not notice this phenomenon for $A_g^1$ and $A_g^2$ consistent with Ref. [5]. (ii) The linewidths of all three phonon modes increase with increasing temperature, while their integrated intensities decrease, consistent with Ref. [5].

## SIV. *Ab initio* Calculations.

Our theoretical approach to investigate CrSBr has been discussed in Ref. [7]. In the density functional theory within the generalized gradient approximation (GGA), magnetic and spin–orbit effects are fully taken into account in our spinor wave functions. The many-body perturbation theory is evaluated within the $GW$ approximation to evaluate the electronic properties. For the optical properties we employ the Bethe-Salpeter equation (BSE).

Our theoretical spectra (see main text) have been calculated with a broadening of 35 meV. For the monolayer we use a mesh of $24 \times 32$ $k$ points, 30 valence and 20 conduction bands. For the bulk BSE calculations, a reduced set of $17 \times 23 \times 3$ k points and $30 \times 4$ bands is explicitly taken into account. It is extrapolated to the converged mesh and band values of the monolayer.

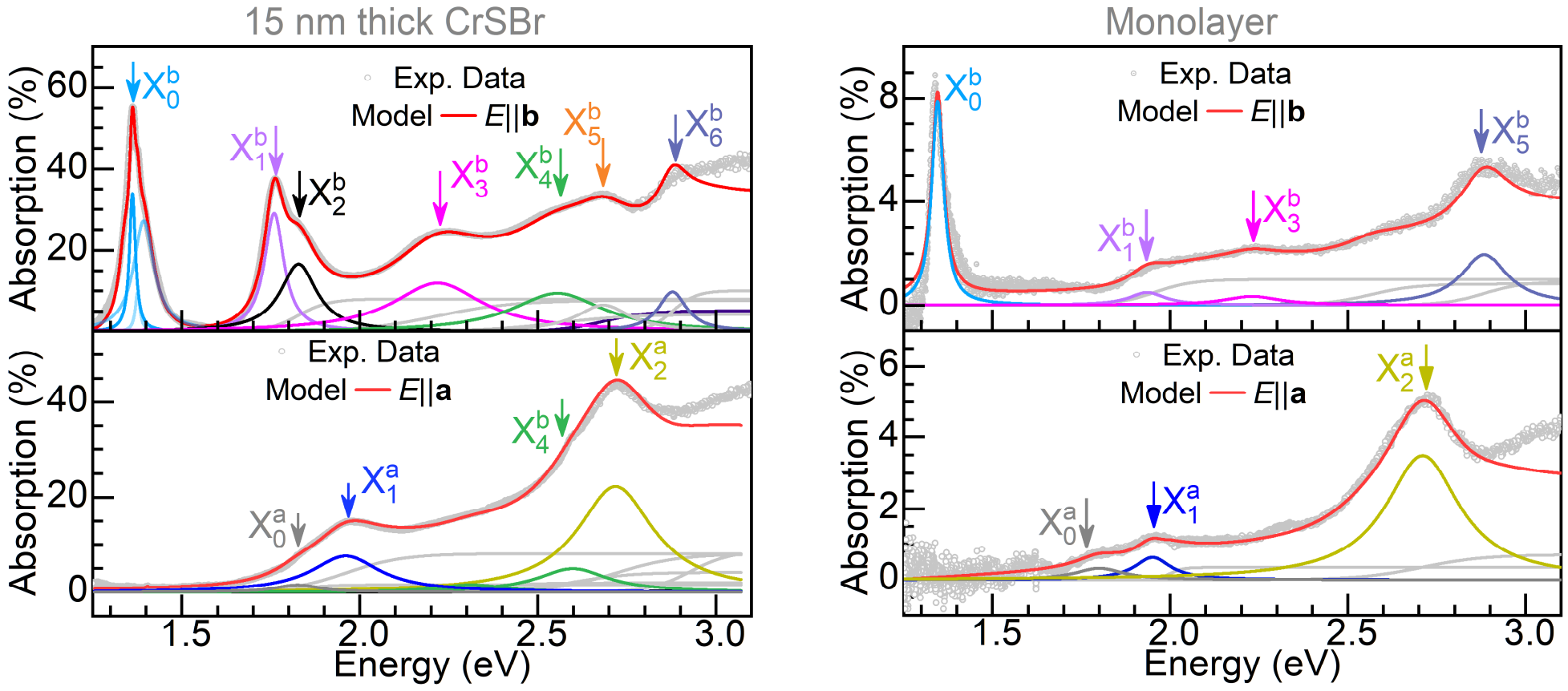

Figure S4. **Line shape modeling.** The experimental absorption spectra with the corresponding fits for both bulk-like and monolayer CrSBr for light polarized along the crystallographic **a-** and **b-**axis. Separate Lorentzian resonance functions for excitonic bound states and Sommerfeld-enhanced step-like functions for the excitonic continuums are shown. Red lines are overall fits for light polarized along both **b-** and **a-**axes.

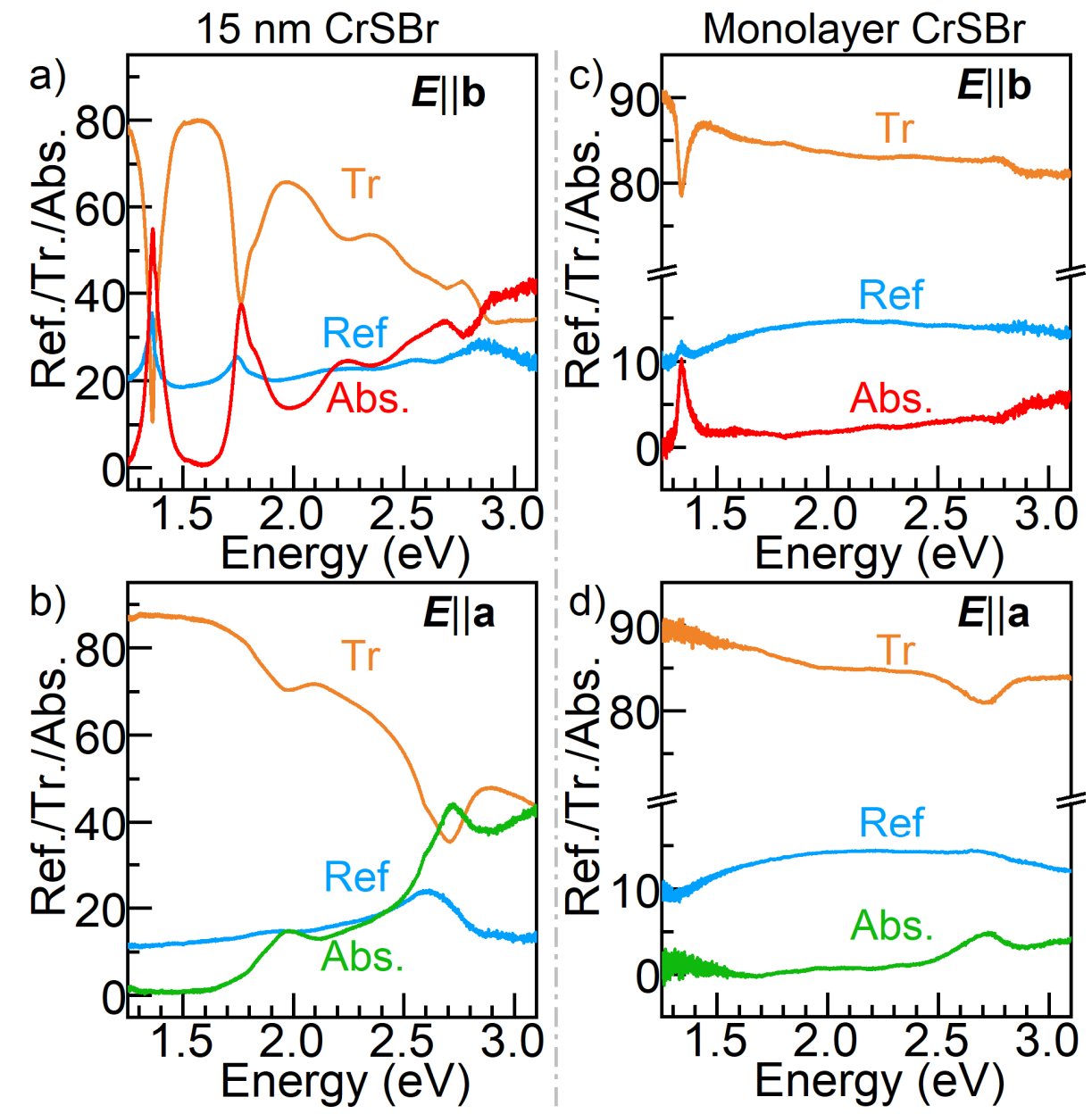

Figure S5. **Polarized optical spectra of CrSBr flakes.** Reflectance (Ref, blue), transmittance (Tr, orange), and absorption (Abs, red/green) spectra measured for a 15 $nm$ thick CrSBr flake with light polarization parallel to the (a) **b**-axis ($\boldsymbol{E} \parallel \mathbf{b}$) and (b) a-axis ($\boldsymbol{E} \parallel \mathbf{a}$). c and d represent corresponding cases for the monolayer. The measurements were performed at $T = 70\ K$.

**References.**